\documentclass[twoside]{article}
\usepackage{waspaa01}
\usepackage{amssymb,amsmath,amsthm,amssymb,commath}
\usepackage[dvipdfm]{hyperref}
\hypersetup{
	bookmarks=true,            
	unicode=false,             
	pdftoolbar=true,           
	pdfmenubar=true,           
	pdffitwindow=false,        
	pdfstartview={FitH},       
	pdftitle={blue},           
	pdfauthor={Author},        
	pdfsubject={Subject},      
	pdfcreator={Creator},      
	pdfproducer={Producer},    
	pdfkeywords={keywords},    
	pdfnewwindow=true,         
	colorlinks=true,           
	linkcolor=blue ,           
	citecolor=blue,            
	filecolor=magenta,         
	urlcolor=cyan              
}
\usepackage{color}
\usepackage{setspace}
\usepackage{caption}
\usepackage{graphicx}
\usepackage{comment}
\usepackage{subfig,float}
\usepackage{tabularx,booktabs}
\usepackage{afterpage}
\usepackage{placeins}

\title{\bfseries A novel nonlinear amplitude-modulation gyroscope incorporating internal resonance}
\twoauthors{}
{S. Amir Mousavi Lajimi, Navid Noori, Amr Marzouk, Behraad Bahreyni, Farid Golnaraghi\\
Mechatronic Systems Engineering, Simon Fraser University,
B.C., Canada V3T 0A3 \\
samirm@sfu.ca, samousav@uwaterloo.ca}
{}{} 
\graphicspath{{Figs/}}
\begin{document}

\maketitle

\section*{\raggedright \small ABSTRACT}

We are presenting the design and the preliminary numerical and experimental analyses of two mismatched Coriolis vibratory gyroscopes incorporating nonlinear modal interaction. 
A novel double-H design includes two clamped-clamped beams and a suspended mass in the middle connected to the base beams via four short cantilevers. Another design is a T-shaped gyro including a primary doubly-clamped beam and a secondary sense beam. 
A combination of analytical, finite element, and experimental analyses are employed to study the characteristics of the nonlinear gyro. 
The drive mode matches the structure's second mode, while the sense mode matches the fundamental mode of the structure. Our preliminary study indicates that the bandwidth and the sensitivity of the rotation rate sensor are improved by employing the nonlinear modal interaction.    
\\
\\
\ {\bfseries\ Keywords: } Gyroscope, rate sensor, internal resonance, sensitivity, bandwidth, noise, saturation

\section *{\raggedright \normalsize  INTRODUCTION}
A conventional Coriolis vibratory gyro (CVG) refers to a micro- or macro-mechanical structure where the Coriolis
effect transfers some energy from a primary oscillation mode to a secondary oscillation mode. The system
used as a sensor, includes a primary (drive) mode excited by an external source and a secondary
(sense) mode activated by the rotation rate-dependent terms. To operate a CVG in open-loop
mode and amplitude-modulation scheme, the amplitude of the sense mode (pickoff) is used
to identify the rotation rate \cite{Curey2004}. Another operation mode for vibratory gyroscopes is the force-to-rebalance, or closed-loop, mode where the sense amplitude is estimated by an external signal proportional to the
rotation rate \cite{Curey2004}. 

The designs of rotation rate sensors (gyroscopes) are evolving to offer higher resolution, larger sensitivity, better performance and a lower fabrication cost. Beam-based gyroscopes offer a relatively simple structure and have been in-use for a long time \cite{ConnorS1983}. Gyroscopes suffer from the low sensitivity and low mechanical scale factor. To overcome these difficulties, researchers have
investigated beam-mass systems both experimentally and analytically \cite{Yang2002,LajimiSPIE2014}. Furthermore, to operate the beam-mass gyroscope using frequency-modulation scheme the closed-form equation relating the angular rotation rate and the modal frequencies has been obtained \cite{LajimiIEEE2014,LajimiIMECE2014} and to exploit nonlinearities for higher performance, the nonlinear response has been characterized \cite{LajimiIDETC2014}.

To increase the sensitivity of an amplitude-modulation gyro, the frequencies of the sense and drive modes are matched \cite{Zaman2008}. However, the bandwidth and other performance parameters are affected adversely and maybe improved by other arrangements of the modes \cite{AcarBook2008}. The other possibility is to use different mode shapes of a structure and operate based on the features of nonlinearly coupled systems to increase the bandwidth and the sensitivity as it has been shown for resonators \cite{Vyas2009}. Furthermore, a parametrically resonating gyroscope offers a higher performance \cite{Ramos2009}. Recently, Golnaraghi et al. \cite{Lajimi2015patapp} revealed an application of internal resonance in designing gyroscopes with larger bandwidths and higher sensitivities \cite{Lajimi2015patapp}.

\begin{figure}[!b!] 
	\begin{center}
		\includegraphics[trim =10mm 38mm 10mm 52mm, clip, width=0.48\textwidth,height=1.7in]{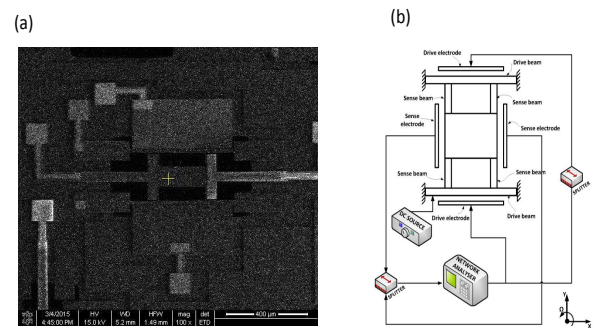}
		\caption{(a) A SEM image of the fabricated double-H shaped gyro and (b) the schematic of the gyroscope and the test setup.}
		\label{fig:MyDesign2}
	\end{center}
\end{figure}

\begin{figure}[!t!] 
	\begin{center}
		\includegraphics[trim =10mm 38mm 10mm 53mm, clip, width=0.49\textwidth,height=1.65in]{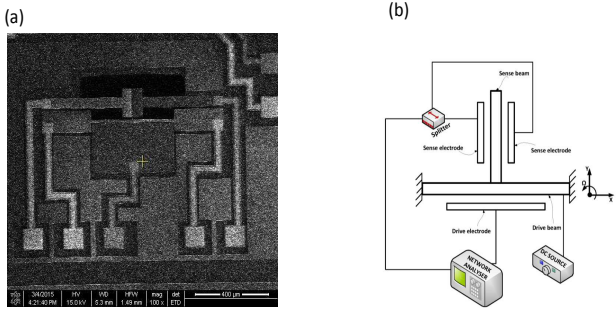}
		\caption{(a) A SEM image of the fabricated T-shaped gyro and (b) the schematic of gyroscope and the test setup.}
		\label{fig:TDesign2}
	\end{center}
\end{figure}

In this work, we propose methods for the design and analysis of an amplitude-modulation gyro which is excited at the frequency of the drive mode being twice the frequency of the sense mode. The first and second frequencies are internally coupled because of the structure's design and offer a higher sensitivity and a larger bandwidth in the sensed response. The finite element analysis is performed in Ansys Mechanical APDL 14.5 package using static, modal, harmonic, and transient analyses. The initial designs are implemented in CoventorWare package and fabricated using a  SOIMUMPs process \cite{Cowen2011}. Two designs are used to study two-to-one internally coupled gyros where the first one, a double-H gyro, is a novel design and the second one, a T-shaped structure, was previously studied as a resonator. A macro-scale model of the T-shaped structure is built and tested on a rate table and the frequency-response curves are obtained. Using a lumped-mass model, a mathematical model of the system is developed and solved numerically to obtain the saturation curves. 
\section *{\raggedright \normalsize DESIGNS and METHODS}
The schematic of the two micro-gyro designs with various configurations of the drive and sense electrodes are provided in Figs. \ref{fig:MyDesign2} and \ref{fig:TDesign2}. In Fig. \ref{fig:MyDesign2}, exciting the base beams along Y-axis, the Coriolis effect induces a response in the X-direction for a single-axis gyro rotating about the Z-axis (out-of-plane). It is worth mentioning that the same structure may be used for in-plane rate measurements placing the drive electrodes along the Z-axis under the drive beam on the Si substrate. The double-H design includes a mass to further amplify the sense signal. 
The second design is a T-shaped structure previously used as a resonator in \cite{Vyas2009}, see Fig. \ref{fig:TDesign2}. 

Both designs are initially analyzed using MEMS CAD software CoventorWare and ANSYS 14.5 Workbench and fabricated using a standard SOIMUMPs process. Scanning electron microscope (SEM) images of the fabricated devices are presented in Figs. \ref{fig:MyDesign2}(a) and \ref{fig:TDesign2}(a) obtained using a FEI Nova NanoSEM 430 SEM System. The structures are designed such that the modal frequency of the second mode is twice the modal frequency of the first mode interacting with each other through nonlinear modal interaction (internal resonance). The pair of sense electrodes provide the required means for differential sensing of the sense displacement. For the double-H gyro, Fig. \ref{fig:MyDesign2}, the drive electrodes operate in 180 degrees phase difference to increase the strength of actuation. Proper instrumentation of the experimentations are shown in Figs. \ref{fig:MyDesign2}(b) and \ref{fig:TDesign2}(b).

\begin{figure}[!h!] 
	\begin{center}
		\includegraphics[trim =55mm 25mm 50mm 15mm, clip, width=0.38\textwidth,height=2in]{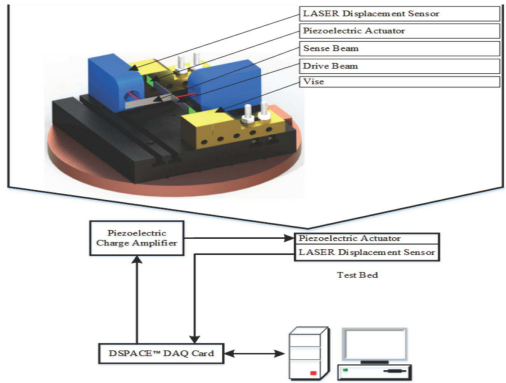}
		\caption{Experimental test setup for the T-shaped gyro. The structure is placed on the table inside the rate table chamber.}
		\label{fig:MacroExp}
	\end{center}
\end{figure}

To properly identify the nonlinear dynamics of the system, a macro-scale model of the T-shaped sensor is built and tested on a rate table, see Fig. \ref{fig:MacroExp}. The parameters of the gyroscope include a base doubly-clamped beam of length 160.20mm, thickness 0.47mm, and width 15mm, and a secondary beam of length 80mm, thickness 0.203mm, and width 15mm. The actuation and sensing mechanisms include two pairs of PSI-5H4E piezoelectric ceramic transducers and a pair of LTS 15/03 and 15/12 laser displacement sensors. The device is mounted on an Ideal Aerosmith 1621-200A-TL rate table and operated and controlled through a customized MATLAB Simulink code and a dSpace data acquisition system. 

\begin{figure}[!h!] 
	\begin{center}
		\includegraphics[trim =10mm 65mm 10mm 55mm, clip, width=0.46\textwidth,height=1in]{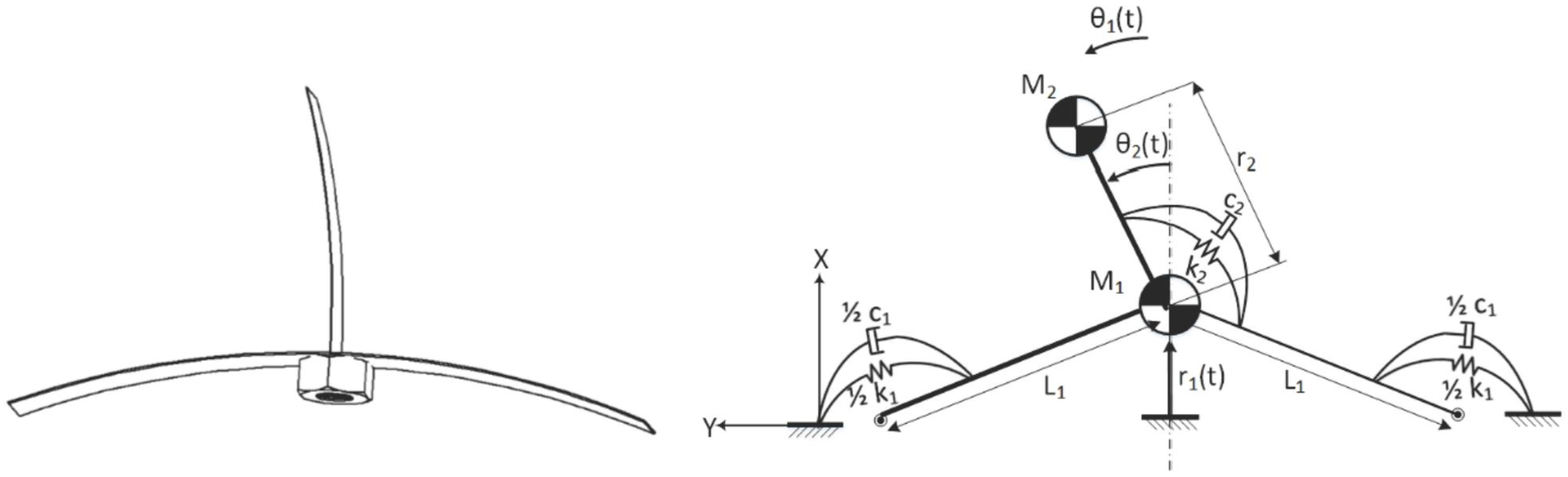}
		\caption{A lumped-mass model of the T-shaped gyroscope.}
		\label{fig:TDesignModel}
	\end{center}
\end{figure}

A simplified mathematical model based on lumped-mass model, Fig. \ref{fig:TDesignModel},  of the T-shaped design, Fig. \ref{fig:TDesign2}, is used to derive the equation of motion in the form
\begin{align}
\label{Eq1}
\begin{bmatrix} M \end{bmatrix}
\begin{Bmatrix} \ddot{r}_1 \\ \ddot{\theta}_2  \end{Bmatrix} + 
\begin{bmatrix} C \end{bmatrix}
\begin{Bmatrix} \dot{r}_1 \\ \dot{\theta}_2  \end{Bmatrix} +
\begin{bmatrix} K \end{bmatrix}
\begin{Bmatrix} K_1 \\ K_2  \end{Bmatrix} =  \begin{Bmatrix} F_{r} + F_e  \\ F_\theta \end{Bmatrix}
\end{align}
where
\begin{align}
\begin{bmatrix} M \end{bmatrix}
 =  \begin{Bmatrix} M_1+M_2 & -M_2 r_2 \sin(\theta_2) \\ -M_2 r_2 \sin(\theta_2) & M_2 r_2^2 \end{Bmatrix} \nonumber
\end{align}
\begin{align}
\begin{bmatrix} C \end{bmatrix}
=  \begin{Bmatrix} \frac{c_1}{L_1^2 - r_1^2} & 0  \\ 0 & c_2 \end{Bmatrix},\,\,
\begin{bmatrix} K \end{bmatrix}
=  \begin{Bmatrix} \frac{\arcsin(\frac{r_1}{L_1})}{L_1\sqrt{1-\frac{r_1^2}{L_1^2}}} & 0  \\ 0 & \theta_2 \end{Bmatrix}\nonumber
\end{align}
\begin{align}
F_r      = &  M_2 r_2 \left(\dot{\theta}_2^2 + \Omega^2 \right) \cos\left(\theta_2\right) + 2 M_2 r_2 \Omega \cos\left(\theta_2\right) \dot{\theta}_2 \nonumber\\
           &  + \left(M_1 + M_2 \right) r_1 \Omega^2 \nonumber \\
F_e  =  &   a_e \sin\left(f_e \, t\right) \nonumber \\
F_\theta =  &  -M_2  r_2 r_1  \Omega^2  \sin(\theta_2) - 2 M_2  r_2 \Omega \cos(\theta_2) \dot{r}_1 \nonumber
\end{align}
where parameters and variables are introduced in Fig. \ref{fig:TDesignModel}. The mathematical model, Eq. \eqref{Eq1}, is used to numerically study the behaviour of the system and to derive a reduced-order model based on a two-variable perturbation method \cite{Golnaraghi1989}. 

\begin{figure}[!h!] 
	\begin{center}
		\includegraphics[trim =10mm 45mm 10mm 55mm, clip, width=0.49\textwidth,height=1.65in]{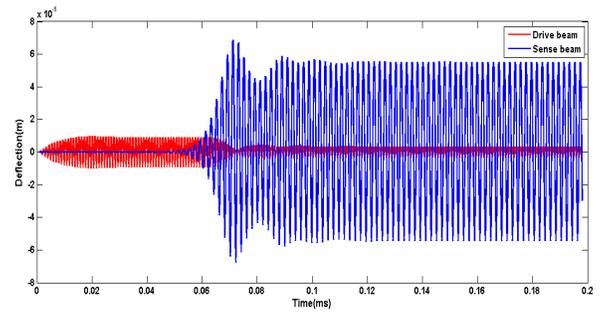}
		\caption{Displacement time histories of the drive and sense beams for micro T-gyro ($\Omega = 10^{\circ}$/s).}
		\label{fig:THAnsysMicroTGyroOmega10}
	\end{center}
\end{figure}

\begin{figure}[!h!] 
	\begin{center}
		\includegraphics[trim =10mm 45mm 15mm 55mm, clip, width=0.49\textwidth,height=1.65in]{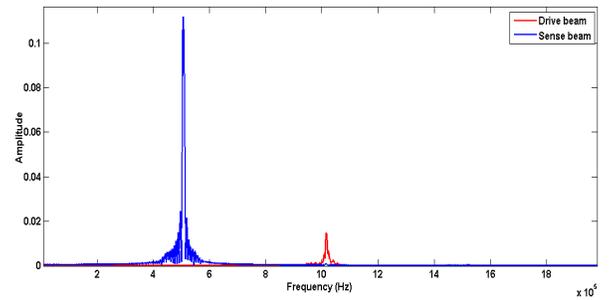}
		\caption{Fast Fourier transforms (FFT) of the displacement time histories for micro T-gyro ($\Omega = 10^{\circ}$/s).}
		\label{fig:FFTAnsysMicroTGyroOmega10}
	\end{center}
\end{figure}

\section *{\raggedright \normalsize RESULTS}
An Ansys Mechanical APDL code is developed to study the static, modal, harmonic, and transient responses of the structures. To perform the finite element analysis of the micro T-shaped gyro in a reasonable amount of time, the quality factor is set to 100 and a two-dimensional model of the structure, Fig. \ref{fig:TDesign2}(a), is studied in Ansys Mechanical using Plane 182 2D structural solid elements. For a T-shaped gyro with a primary 388$\mu$m long clamped-clamped beam with width 20$\mu$m and a secondary 145.5$\mu$m long beam with width 8$\mu$m, the time-histories of the drive and sense displacements are computed using transient analysis and shown in Fig. \ref{fig:THAnsysMicroTGyroOmega10}. The energy is transferred from the drive beam (mode) to the sense beam (mode) through internal resonance and the Coriolis effect giving rise to the large amplitude of the sense beam. A fast Fourier transform (FFT) of the sense and drive time-histories are given in Fig. \ref{fig:FFTAnsysMicroTGyroOmega10} indicating the frequency components of the time-responses. The excitation frequency (the second peak) in the drive direction transforms into the primary frequency (the first peak) in the sense direction through the internal resonance and the Coriolis effect.

\begin{figure}[!h!] 
	\begin{center}
		\includegraphics[trim =10mm 20mm 15mm 45mm, clip, width=0.46\textwidth,height=1.7in]{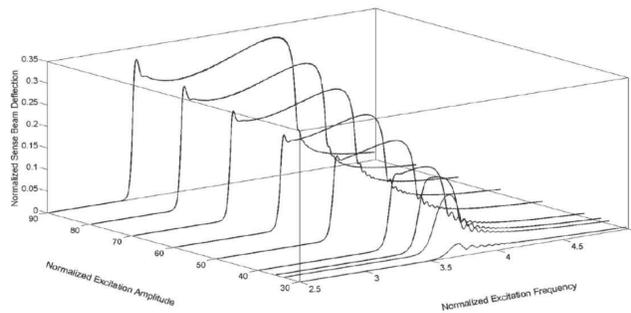}
		\caption{Experimental frequency-response curves of the sense beam for macro T-gyro showing an increase in the bandwidth of the response.}
		\label{fig:ExpFRCSense}
	\end{center}
\end{figure}

In Fig. \ref{fig:ExpFRCSense}, the experimental frequency-response curves for macro-structure of Fig. \ref{fig:MacroExp} are presented. Repeating the experiment for a set of excitation amplitudes, a family of frequency-response curves are obtained showing an increase in the bandwidth and the sense amplitude. For small excitation amplitudes, the response is similar to linear systems, however for large excitation amplitudes a flat region appears in the frequency-response curves offering a higher stability and a lesser sensitivity to where the measurements are taken along the frequency-response curve. 

\begin{figure}[!h!] 
	\begin{center}
		\includegraphics[trim =5mm 10mm 5mm 5mm, clip, width=0.46\textwidth,height=1.7in]{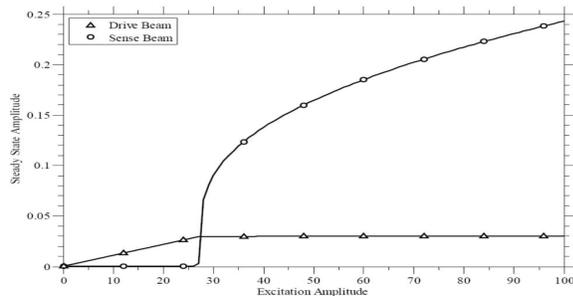}
		\caption{Numerical force-response curves of the sense and drive modes for macro T-gyro showing the nonlinear saturation of the drive amplitude.}
		\label{fig:NumForRC}
	\end{center}
\end{figure}

A numerical solution of the equation of motion, Eq. \ref{Eq1}, is computed for an increasing amplitude of excitation by direct integration and plotted in Fig. \ref{fig:NumForRC}. The sense mode starts to grow after a certain threshold of the excitation amplitude is passed where the drive amplitude reaches to a limit and remains unchanged. The saturation phenomena appears in systems having internally coupled modes of vibration. Similar results are obtained using a two-variable perturbation method and experimental analysis. 

%
%

\section*{\raggedright \normalsize CONCLUSIONS}
%
The coupled-system results in a multi-frequency response in the sense direction of the proposed novel Coriolis vibratory gyros. Initial results show improvements in the sensitivity and the bandwidth of the open-loop gyroscope. Using internal resonance offers a method to overcome the complexities of a closed-loop gyro. Employing the two-to-one internal resonance phenomenon, a wide bandwidth high amplitude response in the sense direction is achieved improving the stability and the sensitivity of the device. Furthermore, separating the drive and sense mode frequencies (for example the drive frequency is twice bigger than the sense frequency-{\it i.e} a 2:1 internal resonance), the sensed signal could be filtered for frequencies up to near the frequency of the drive mode reducing the noise effects around the natural frequency of the sense mode. It is expected to see a considerable improvement in the long term stability of the sensor by reducing the effects of the electronic noise on the readout circuit. Further investigations are underway to properly characterize the system response, examine the possibility of using the method to improve the performance of the amplitude-modulation gyroscopes, and to examine the possibility of using a closed-loop to control and manage the effects of nonlinearities.


\renewcommand{\refname}{\leftline{ \bfseries \normalsize REFERENCES}}
%
%
%
%
%
\bibliographystyle{asmems4}
\bibliography{RefsList}
\end{document}